\documentclass[reprint,amsmath,superscriptaddress,amssymb,aps, physrev]{revtex4-2}
\usepackage[T1]{fontenc}
\usepackage{mathptmx}
\usepackage{xcolor}
\usepackage{textcomp}
\usepackage[colorlinks=true,allcolors=blue]{hyperref}
\usepackage{dblfloatfix}
\usepackage{siunitx}
\usepackage{braket}
\usepackage{graphicx}
\usepackage{dcolumn}
\usepackage{bm}

\usepackage{tikz}
\definecolor{lime}{HTML}{A6CE39}
\DeclareRobustCommand{\orcidicon}{
	\begin{tikzpicture}
	\draw[lime, fill=lime] (0,0) 
	circle [radius=0.16] 
	node[white] {{\fontfamily{qag}\selectfont \tiny ID}};
	\draw[white, fill=white] (-0.0625,0.095) 
	circle [radius=0.007];
	\end{tikzpicture}
	\hspace{-2mm}
}
\foreach \x in {A, ..., Z}{\expandafter\xdef\csname orcid\x\endcsname{\noexpand\href{https://orcid.org/\csname orcidauthor\x\endcsname}
			{\noexpand\orcidicon}}
}
\foreach \x in {A, ..., Z}{\expandafter\xdef\csname orcids\x\endcsname{\noexpand\href{https://orcid.org/\csname orcidsauthor\x\endcsname}
			{\noexpand\orcidicon}}
}

\begin{document}

\preprint{APS/123-QED}

\title{\textbf{Ultracold neutron energy spectrum and storage properties from magnetically induced spin depolarization} 
}%

\author{N.\,J.~Ayres}
\affiliation{
Institute for Particle Physics and Astrophysics,
ETH Z\"urich,
CH-8093 Zurich, Switzerland}

\author{G.~Ban}
\affiliation{Normandie Universit\'e, ENSICAEN, UNICAEN, CNRS/IN2P3, LPC Caen, 14000 Caen, France}

\author{G.~Bison\orcidK{}}
\affiliation{Paul Scherrer Institut, CH-5232 Villigen PSI, Switzerland}

\author{K.~Bodek}
\affiliation{Marian Smoluchowski Institute of Physics, Jagiellonian University, 30-348 Cracow, Poland}

\author{V.~Bondar}
\affiliation{
Institute for Particle Physics and Astrophysics,
ETH Z\"urich,
CH-8093 Zurich, Switzerland}

\author{T.~Bouillaud}
\affiliation{Universit\'e Grenoble Alpes, CNRS, Grenoble INP, LPSC-IN2P3, 38026 Grenoble, France}

\author{D.~Bowles}
\affiliation{Department of Physics and Astronomy, University of Kentucky, Lexington 40506, Kentucky, USA}

\author{G.\ L.~Caratsch}
\affiliation{
Institute for Particle Physics and Astrophysics,
ETH Z\"urich,
CH-8093 Zurich, Switzerland}
\affiliation{Paul Scherrer Institut, CH-5232 Villigen PSI, Switzerland}

\author{E.~Chanel}
\affiliation{Laboratory for High Energy Physics and
Albert Einstein Center for Fundamental Physics,
University of Bern, CH-3012 Bern, Switzerland}
\affiliation{Present address: Institut Laue-Langevin - 71 avenue des Martyrs CS 20156, 38042 GRENOBLE Cedex 9 - France}
\author{W.~Chen\orcidU{}}
\affiliation{
Institute for Particle Physics and Astrophysics,
ETH Z\"urich,
CH-8093 Zurich, Switzerland}
\affiliation{Paul Scherrer Institut, CH-5232 Villigen PSI, Switzerland}

\author{P.-J.~Chiu\orcidE{}}
\affiliation{
Institute for Particle Physics and Astrophysics,
ETH Z\"urich,
CH-8093 Zurich, Switzerland}
\affiliation{Paul Scherrer Institut, CH-5232 Villigen PSI, Switzerland}
\affiliation{Present address: National Taiwan University, 106319 Taipei, Taiwan}

\author{C.\,B.~Crawford}
\affiliation{Department of Physics and Astronomy, University of Kentucky, Lexington 40506, Kentucky, USA}

\author{V.~Czamler}
\affiliation{Universit\'e Grenoble Alpes, CNRS, Grenoble INP, LPSC-IN2P3, 38026 Grenoble, France}

\author{M.~Daum}
\affiliation{Paul Scherrer Institut, CH-5232 Villigen PSI, Switzerland}

\author{C.\,B.~Doorenbos}
\affiliation{
Institute for Particle Physics and Astrophysics,
ETH Z\"urich,
CH-8093 Zurich, Switzerland}
\affiliation{Paul Scherrer Institut, CH-5232 Villigen PSI, Switzerland}

\author{M.~Ferry\orcidY{}}
\affiliation{Universit\'e Grenoble Alpes, CNRS, Grenoble INP, LPSC-IN2P3, 38026 Grenoble, France}

\author{M.~Fertl\orcidM{}}
\affiliation{Institute of Physics, Johannes Gutenberg University, D-55128 Mainz, Germany}

\author{A.~Fratangelo\orcidsB{}}
\affiliation{Laboratory for High Energy Physics and
Albert Einstein Center for Fundamental Physics,
University of Bern, CH-3012 Bern, Switzerland}

\author{D.~Galbinski\orcidN{}}
\affiliation{Normandie Universit\'e, ENSICAEN, UNICAEN, CNRS/IN2P3, LPC Caen, 14000 Caen, France}

\author{W.\,C.~Griffith\orcidS{}}
\affiliation{Department of Physics and Astronomy, University of Sussex, Falmer, Brighton BN1 9QH, United Kingdom}

\author{Z.\,D.~Grujic}
\affiliation{Institute of Physics, Photonics Center, University of Belgrade, 11080 Belgrade, Serbia}

\author{K.~Kirch\orcidB{}}
\affiliation{
Institute for Particle Physics and Astrophysics,
ETH Z\"urich,
CH-8093 Zurich, Switzerland}
\affiliation{Paul Scherrer Institut, CH-5232 Villigen PSI, Switzerland}

\author{V.~Kletzl\orcidF{}}
\affiliation{
Institute for Particle Physics and Astrophysics,
ETH Z\"urich,
CH-8093 Zurich, Switzerland}
\affiliation{Paul Scherrer Institut, CH-5232 Villigen PSI, Switzerland}
\affiliation{Present address: Marietta Blau Institute for Particle Physics, Austrian Academy of Sciences, 1010 Vienna, Austria}

\author{B.~Lauss\orcidG{}}
\affiliation{Paul Scherrer Institut, CH-5232 Villigen PSI, Switzerland}

\author{T.~Lefort\orcidZ{}}
\affiliation{Normandie Universit\'e, ENSICAEN, UNICAEN, CNRS/IN2P3, LPC Caen, 14000 Caen, France}

\author{A.~Lejuez}
\affiliation{Normandie Universit\'e, ENSICAEN, UNICAEN, CNRS/IN2P3, LPC Caen, 14000 Caen, France}

\author{R.~Li\orcidsA{}}
\affiliation{Instituut voor Kern- en Stralingsfysica, University of Leuven, B-3001 Leuven, Belgium}

\author{K.~Michielsen\orcidW{}}
\affiliation{Universit\'e Grenoble Alpes, CNRS, Grenoble INP, LPSC-IN2P3, 38026 Grenoble, France}

\author{J.~Micko}
\affiliation{Laboratory for High Energy Physics and
Albert Einstein Center for Fundamental Physics,
University of Bern, CH-3012 Bern, Switzerland}

\author{P.~Mullan\orcidL{}}
\affiliation{
Institute for Particle Physics and Astrophysics,
ETH Z\"urich,
CH-8093 Zurich, Switzerland}

\author{A.~Mullins}
\affiliation{Department of Physics and Astronomy, University of Kentucky, Lexington 40506, Kentucky, USA}

\author{O.~Naviliat-Cuncic}
\affiliation{Normandie Universit\'e, ENSICAEN, UNICAEN, CNRS/IN2P3, LPC Caen, 14000 Caen, France}

\author{D.~Pais}
\affiliation{
Institute for Particle Physics and Astrophysics,
ETH Z\"urich,
CH-8093 Zurich, Switzerland}
\affiliation{Paul Scherrer Institut, CH-5232 Villigen PSI, Switzerland}

\author{F.\,M.~Piegsa\orcidO{}}
\affiliation{Laboratory for High Energy Physics and
Albert Einstein Center for Fundamental Physics,
University of Bern, CH-3012 Bern, Switzerland}

\author{G.~Pignol\orcidJ{}}
\affiliation{Universit\'e Grenoble Alpes, CNRS, Grenoble INP, LPSC-IN2P3, 38026 Grenoble, France}

\author{C.~Pistillo}
\affiliation{Laboratory for High Energy Physics and
Albert Einstein Center for Fundamental Physics,
University of Bern, CH-3012 Bern, Switzerland}

\author{D.~Rebreyend}
\affiliation{Universit\'e Grenoble Alpes, CNRS, Grenoble INP, LPSC-IN2P3, 38026 Grenoble, France}

\author{I.~Rien\"acker\orcidX{}}
\affiliation{
Institute for Particle Physics and Astrophysics,
ETH Z\"urich,
CH-8093 Zurich, Switzerland}
\affiliation{Paul Scherrer Institut, CH-5232 Villigen PSI, Switzerland}

\author{D.~Ries}
\affiliation{Paul Scherrer Institut, CH-5232 Villigen PSI, Switzerland}

\author{S.~Roccia\orcidR{}}
\affiliation{Universit\'e Grenoble Alpes, CNRS, Grenoble INP, LPSC-IN2P3, 38026 Grenoble, France}

\author{D.~Rozpedzik}
\affiliation{Marian Smoluchowski Institute of Physics, Jagiellonian University, 30-348 Cracow, Poland}

\author{W.~Saenz-Arevalo}
\affiliation{Normandie Universit\'e, ENSICAEN, UNICAEN, CNRS/IN2P3, LPC Caen, 14000 Caen, France}

\author{L.~Sanchez-Real~Zielniewicz\orcidP{}}
\affiliation{
Institute for Particle Physics and Astrophysics,
ETH Z\"urich,
CH-8093 Zurich, Switzerland}

\author{P.~Schmidt-Wellenburg\orcidC{}}
\affiliation{Paul Scherrer Institut, CH-5232 Villigen PSI, Switzerland}

\author{E.\ P.~Segarra\orcidA{}}
\thanks{\bf Corresponding author: efrain.segarra@psi.ch}
\affiliation{Paul Scherrer Institut, CH-5232 Villigen PSI, Switzerland}
\affiliation{
Institute for Particle Physics and Astrophysics,
ETH Z\"urich,
CH-8093 Zurich, Switzerland}

\author{L.~Segner\orcidQ{}}
\affiliation{
Institute for Particle Physics and Astrophysics,
ETH Z\"urich,
CH-8093 Zurich, Switzerland}

\author{N.~Severijns\orcidH{}}
\affiliation{Instituut voor Kern- en Stralingsfysica, University of Leuven, B-3001 Leuven, Belgium}

\author{K.~Svirina\orcidV{}}
\affiliation{Universit\'e Grenoble Alpes, CNRS, Grenoble INP, LPSC-IN2P3, 38026 Grenoble, France}
\affiliation{Present address: Institut Laue–Langevin, CS 20156, 38042 Grenoble Cedex 9, France and Physikalisches Institut, Universitat Heidelberg, Im Neuenheimer Feld 226, 69120 Heidelberg, Germany
}

\author{K.\,S.~Tanaka\orcidT{}}
\affiliation{Paul Scherrer Institut, CH-5232 Villigen PSI, Switzerland}
\affiliation{Present address: Waseda University, Tokyo 169-8555, Japan}

\author{J.~Thorne}
\affiliation{Laboratory for High Energy Physics and
Albert Einstein Center for Fundamental Physics,
University of Bern, CH-3012 Bern, Switzerland}
\affiliation{
Institute for Particle Physics and Astrophysics,
ETH Z\"urich,
CH-8093 Zurich, Switzerland}

\author{J.~Vankeirsbilck}
\affiliation{Instituut voor Kern- en Stralingsfysica, University of Leuven, B-3001 Leuven, Belgium}

\author{N.~von Schickh}
\affiliation{
Institute for Particle Physics and Astrophysics,
ETH Z\"urich,
CH-8093 Zurich, Switzerland}
\affiliation{Paul Scherrer Institut, CH-5232 Villigen PSI, Switzerland}

\author{N.~Yazdandoost\orcidD{}}
\affiliation{Paul Scherrer Institut, CH-5232 Villigen PSI, Switzerland}


\author{J.~Zejma}
\affiliation{Marian Smoluchowski Institute of Physics, Jagiellonian University, 30-348 Cracow, Poland}

\author{N.~Ziehl\orcidI{}}
\affiliation{
Institute for Particle Physics and Astrophysics,
ETH Z\"urich,
CH-8093 Zurich, Switzerland}

\author{G.~Zsigmond}
\affiliation{Paul Scherrer Institut, CH-5232 Villigen PSI, Switzerland}

\collaboration{The nEDM collaboration at PSI}

\date{\today}

\begin{abstract}
We present a novel method for extracting the energy spectrum of ultracold neutrons from magnetically induced spin depolarization measurements using the n2EDM apparatus. This method is also sensitive to the storage properties of the materials used to trap ultracold neutrons, specifically, whether collisions are specular or diffuse. 
We highlight the sensitivity of this new technique by comparing the two different storage chambers of the n2EDM experiment.
We validate the extraction by comparing to an independent measurement for how this energy spectrum is polarized through a magnetic-filter, and finally, we calculate the neutron center-of-mass offset, an important systematic effect for measurements of the neutron electric dipole moment.
\end{abstract}

\maketitle



\section{\label{sec:introduction}Introduction}
Ultracold neutrons (UCNs) are free neutrons of very low-energy (less than $\sim300$~\si{\nano\electronvolt}) and can be materially, magnetically, or gravitationally trapped for measurements on the same timescale as the neutron lifetime~\cite{Ignatovich,Golub1991,Steyerl,HFSTrapPSI2011,Ezhov,UCNtau,Tauspect2024}.
This enables fundamental precision physics experiments to take place, such as studying the neutron lifetime and, as discussed here, the neutron electric dipole moment (nEDM) with the n2EDM apparatus~\cite{n2edmDesign2021}.

In many of these experiments, a significant limitation in assessing systematic effects is the poor knowledge of the UCN energy spectrum in the trap. 
In neutron lifetime experiments, UCNs that are marginally trapped can mimic shorter neutron lifetime~\cite{NIH,UCNtau}, and thus, knowing the energy spectrum within the trap can allow better control of systematics. In storage nEDM experiments that utilize thermal atoms for co-magnetometry~\cite{GreenHg}, the vertical center-of-mass offset of the slow UCN ensemble and its ``sagging'' in the trap (with respect to the geometrical center) can lead to a sizable difference in the sensed magnetic field between UCN and atoms in the presence of a magnetic field gradient (thereby leading to resonant frequency shifts)~\cite{GravDepolarization_Pendlebury,GravDepolar2015,n2edmDesign2021}. Additionally for nEDM experiments, the motion of particles in a non-uniform magnetic field with an electric field induces a ``false'' EDM signal proportional to the squared velocity~\cite{GeometricPhase_Pendlebury,DipoleFieldGeometericPhase,FalsenEDMHg,MagFieldUni2019}, and thus precise knowledge of the energy spectrum will be important to estimate this systematic effect.

Beyond the energy spectrum of UCNs in the trap, 
the specularity of reflections and loss-per-bounce of the material trap play substantial roles. The roughness of the trap surfaces determines the fraction of specular reflections, which influences the spin depolarization~\cite{MCUCN,MagFieldUni2019} and influences the characteristic times to fill and empty the trap volume.

Here we present a new method for extracting the UCN energy spectrum and the fraction of specular reflection of material traps using magnetically induced spin depolarization. Unlike other approaches to access the energy spectrum~\cite{PSI_TOF,SUN2_TOF}, which often require experimental modifications (such as changing storage materials or detector heights) or suffer from extrapolating measurements to the original setup, this work relies solely on magnetic gradients that are generated from coils surrounding the material trap. While similar to  Refs.~\cite{GravDepolar2015,GravDepolar2015_2} in extracting the spectral information, here we can simultaneously extract the fraction of specular reflection of the material trap for the first time. We also compare our extracted energy spectrum to an independent measurement of how the spectrum is polarized through a magnetic-filter, and we provide precise predictions for the vertical center-of-mass offset from the energy spectrum, both important inputs for the n2EDM experiment.


\section{\label{sec:mag-depolarization}Magnetic spin depolarization}
In the n2EDM experiment, UCNs are spin-polarized and then filled into a material trap positioned inside vertical, collinear electric and magnetic fields (also collinear with gravity). The resonance frequency in the $940$~\si{\nano\tesla} magnetic field and $15$~\si{\kilo\volt\per\centi\meter} electric field is then extracted via Ramsey's method of separated oscillating fields~\cite{Ramsey1950}, whereby a magnetic pulse rotates the spins into a coherent superposition in the transverse plane, they freely precess for some time $T$, and then a second pulse converts the accumulated phase into a spin-population difference. 

During this process, UCN spins can be depolarized through several channels. Magnetically-induced depolarization can occur through \textit{intrinsic} depolarization, due to magnetic field inhomogeneities, and through an energy-dependent relative dephasing of different UCN energy groups in a vertical magnetic-field gradient, referred to as \textit{gravitationally enhanced depolarization}. 

\begin{figure}[tp]
    \centering
    \includegraphics[width=0.95\columnwidth]{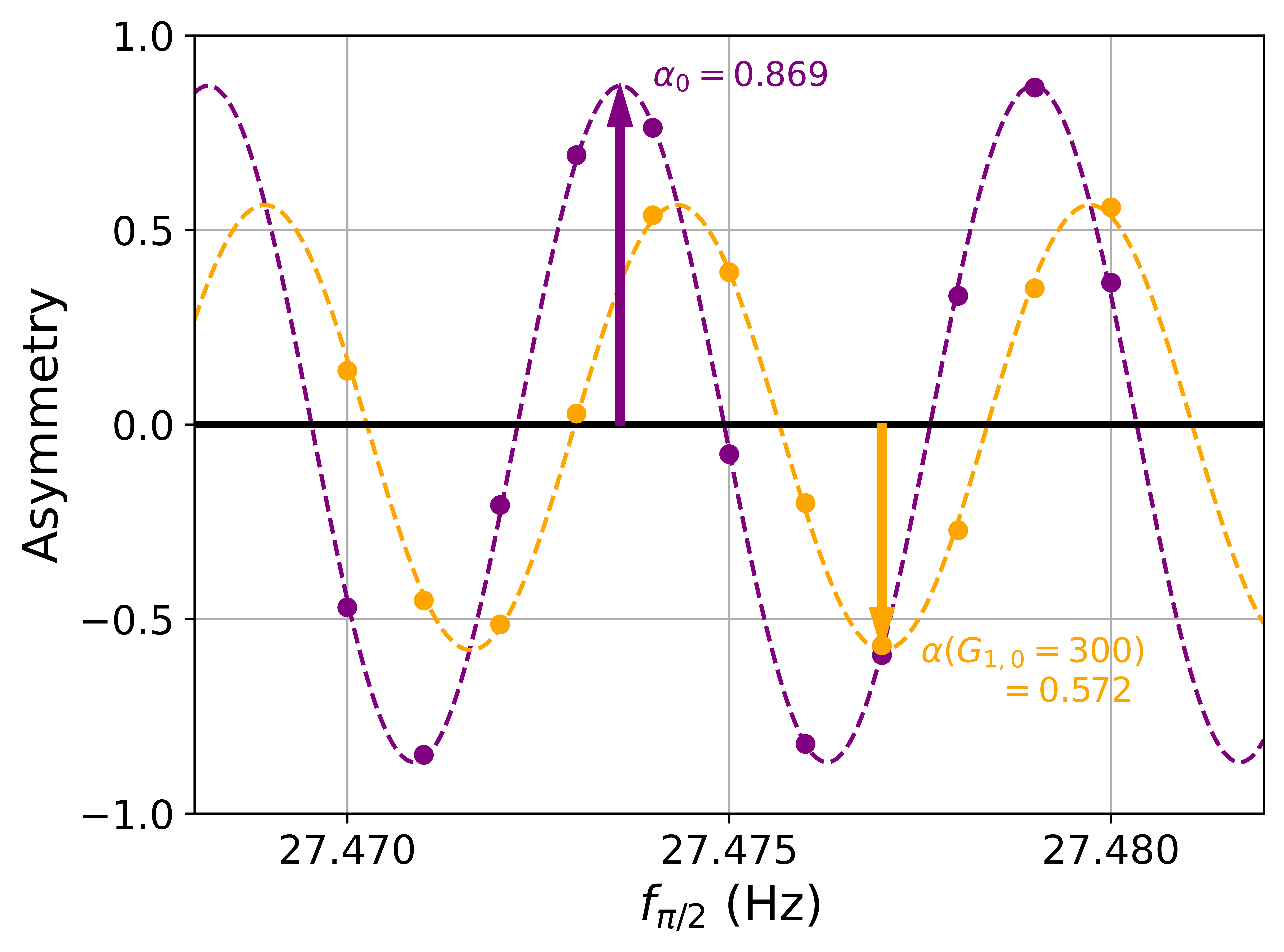}
    \caption{Measured spin asymmetry $A=(N_\uparrow-N_\downarrow)/(N_\uparrow+N_\downarrow)$ as a function of the driving frequency to induce a $\pi/2$ spin flip pulse, $f_{\pi/2}$, in two magnetic field configurations. (Purple): $B=940$~\si{\nano\tesla} with no vertical gradient. (Orange): $B=940$~\si{\nano\tesla} with a vertical gradient of $G_{1,0}=300$~\si{\pico\tesla\per\centi\meter}. Following Eq.~\ref{eqn:grav_depolarization}, $\alpha$ decreases with $G$. The statistical uncertainty on the asymmetry is shown but not visible and on the order of $1\%$. In both cases, the precession time is $T=180$~\si{\second}. The asymmetry is fitted to $A(f_{\pi/2})=-\alpha \cos{\left( \frac{\pi}{\Delta \nu}\left( f_{\pi/2}-f_n\right) \right)}$ in order to extract $\alpha$, where $1/\Delta \nu=2T+8t_{\pi/2}/\pi$ and the duration of the $\pi/2$ pulse is set to $t_{\pi/2}=2$~\si{\second}. The curves shift in $f_{\pi/2}$ due to an induced frequency shift in $f_n$ proportional to $G$.}
    \label{fig:visibility}
\end{figure}

These two channels of depolarization have been thoroughly discussed in Refs.~\cite{GravDepolar2015,GravDepolar2015_2,MagFieldUni2019,n2edmDesign2021}, and can be summarized by the following equations. Gravitationally enhanced depolarization is expressed as
\begin{equation}
    \alpha_\mathrm{grav}(G_{1,0}) = \alpha_0 \int n(\epsilon) \cos{\Big( \gamma_n G_{1,0} T \left(\bar{z}(\epsilon) - \braket{z}\right)  \Big)} d\epsilon, 
    \label{eqn:grav_depolarization}
\end{equation}
where $n(\epsilon)$ is the energy spectrum, $\alpha_0$ is the visibility at zero gradient (see Figure~\ref{fig:visibility}), $T=180$~\si{\second} is the spin precession time, $\gamma_n$ is the neutron gyromagnetic ratio, $G_{1,0}$ is an applied vertical gradient. The vertical center-of-mass offset of the neutron ensemble is $\braket{z}$, where $\bar{z}(\epsilon)$ is the offset for each energy group. See Refs.~\cite{GravDepolar2015,GravDepolar2015_2,n2edmDesign2021} for discussion on $\bar{z}(\epsilon)$. 

Intrinsic depolarization is characterized by the depolarization rate $T_{2,\mathrm{mag}}$, and defined as,
\begin{align}
    \alpha_\mathrm{int}(G_i)=&\ \alpha_0\int n(\epsilon) \exp{\left(-T/T_{2,\mathrm{mag}}
    \label{eqn:tau2mag}
    (\epsilon,G_i)\right)}  d\epsilon  \\
    {1/T_{2,\mathrm{mag}}(\epsilon,G_i)} =&\ \gamma_n^2 \int \braket{b_{z,i}(0) b_{z,i}(\tau)} d\tau \equiv \gamma_n^2 \braket{b_{z,i}^2} \tau_i(\epsilon). \nonumber
\end{align}
Here, $\braket{b_{z,i}^2}$ is the volume-averaged longitudinal magnetic perturbation. The correlation time $\tau_i(\epsilon)$ encodes properties of the material trap, such as the diffuse reflection probability of the surfaces (discussed below)~\cite{n2edmDesign2021}. The exact depolarization rate and correlation time depends on the magnetic pertubation created from the gradient $G_i$. Using the harmonic field description in Ref.~\cite{MagFieldUni2019}, a linear horizontal gradient $G_{1,1}=\delta B_z/\delta x$ would create a perturbation $\braket{b_{z,[1,1]}^2} = G_{1,1}^2 {R^2}/{4}$ in a cylindrical trap of radius $R$. 

The correlation time, $\tau_i(\epsilon)$, is related to the autocorrelation of the longitudinal disturbance, $\braket{b_{z,i}(t)b_{z,i}(t+\tau)}$, and therefore depends on the directionality of the magnetic perturbation $b_{z,i}$. To calculate the correlation functions, energy-dependent trajectories of UCNs in the trap are required. This strongly depends on how specular or diffuse the UCN scattering is on the surfaces of the trap, as this will drastically change the correlation time for different field gradients: the specularity can determine how long a UCN can stay on a given trajectory without exploring the full volume of the storage chamber and field gradient. To model this, we use TOMAt~\cite{TOMAT} to simulate trajectories $x(t), y(t), z(t)$ in the absence of gravity. 
Within TOMAt, we specify the volume of the trap and the diffuse properties of the storage walls. For n2EDM, these are the diffuseness of the electrodes (the floor and ceiling), $p_\mathrm{electrode}$, and the insulator rings $p_\mathrm{ring}$ (the cylindrical walls). Using TOMAt, we calculate a grid of $\tau_{1,1}(p_\mathrm{electrode},p_\mathrm{ring})$ (for a perturbation $G_{1,1}$) and then perform an interpolation (similarly for $\tau_{2,0}$). Here, $p=0$ is perfectly specular reflections and $p=1$ is perfectly diffuse. The correlation times for a given diffuseness were also compared to MCUCN~\cite{MCUCN,MCUCN_2} and starUCN~\cite{starUCN} including gravity. We achieve consistent results, see online supplemental materials.

\section{\label{sec:depolarization-data}Energy spectrum and storage properties extraction}

The work described here was performed using the n2EDM apparatus~\cite{n2edmDesign2021}. n2EDM uses two vertically-stacked, cylindrical storage chambers with DLC-coated electrodes and quartz insulator rings. Each chamber is $12$~\si{\centi\meter} in height and has a diameter of $80$~\si{\centi\meter}. As the UCN are filled into the two chambers differently, they have access to different UCN energy spectra. The top chamber is filled from the top, while the bottom chamber is filled from the bottom. In this work, the total energy $\epsilon$ is taken with respect to the floor of each chamber separately. Due to the filling of the top chamber, the minimum energy in the top chamber is approximately $\sim18$~\si{\nano\electronvolt}. 

Surrounding the two storage chambers is a set of coils that can generate magnetic field gradients, such as vertical gradients $G_{1,0}$ or quadratic gradients $G_{2,0}$. 
The UCNs are measured by a spin-sensitive detector scheme~\cite{USSA,n2edmDesign2021} to construct the spin-asymmetry $(N_\uparrow-N_\downarrow)/(N_\uparrow+N_\downarrow)$. For a given gradient $G_i$, we measure the spin-asymmetry as a function of the Ramsey-pulse drive frequency, and then fit it to extract $\alpha(G_i)$, as shown in Figure~\ref{fig:visibility}. 
While the free precession time was fixed to $T=180$~\si{\second} in each data set, the overall neutron storage time, from closing the trap to opening it again, was $200$~\si{\second}. This is to accommodate the spin manipulation and provide for additional buffer time. The longitudinal depolarization in n2EDM is negligible over these additional $20$~\si{\second} (see online supplemental materials).

\begin{figure}[th]
\centering
\includegraphics[width=0.9\columnwidth]{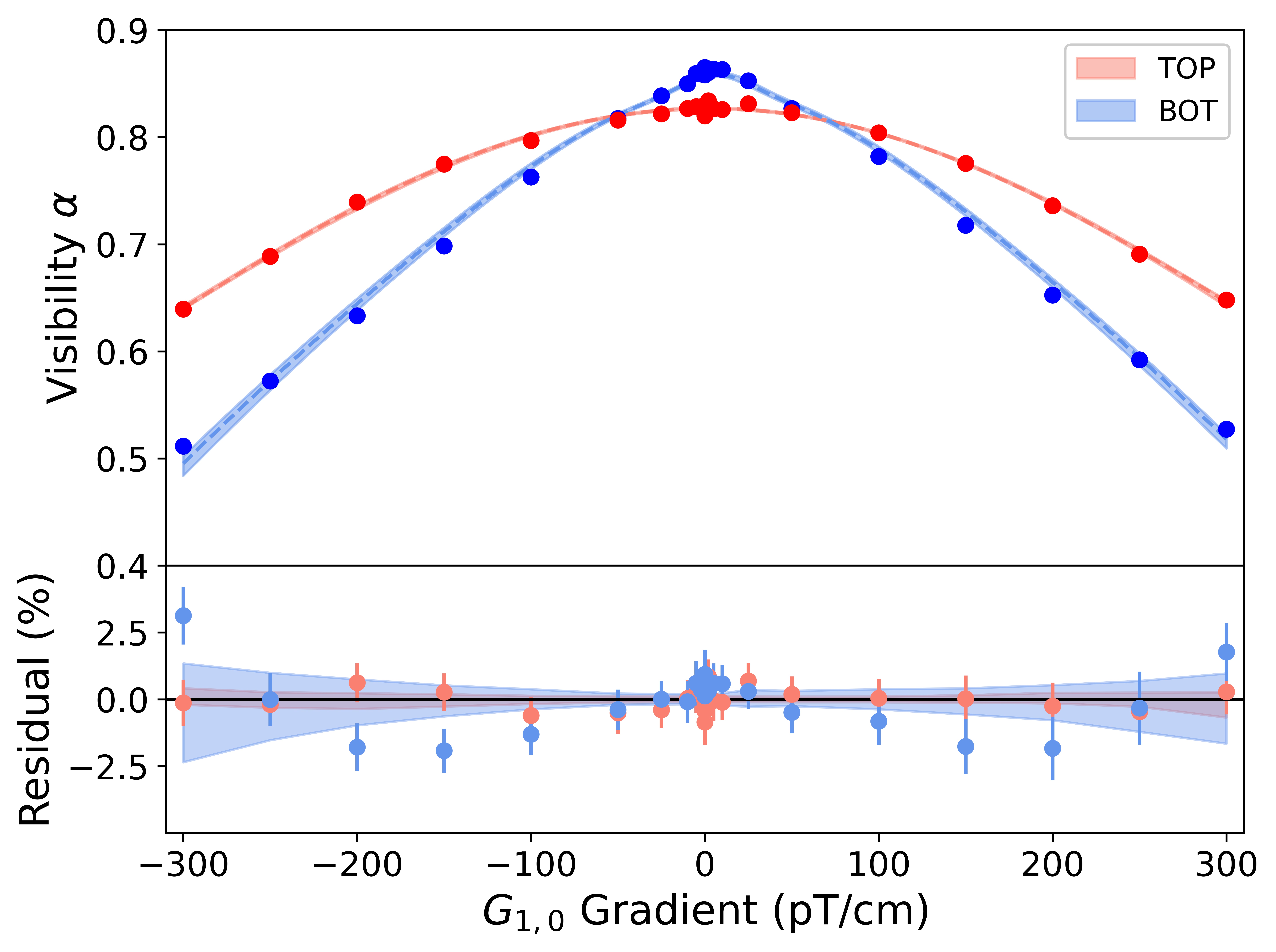}
\includegraphics[width=0.9\columnwidth]{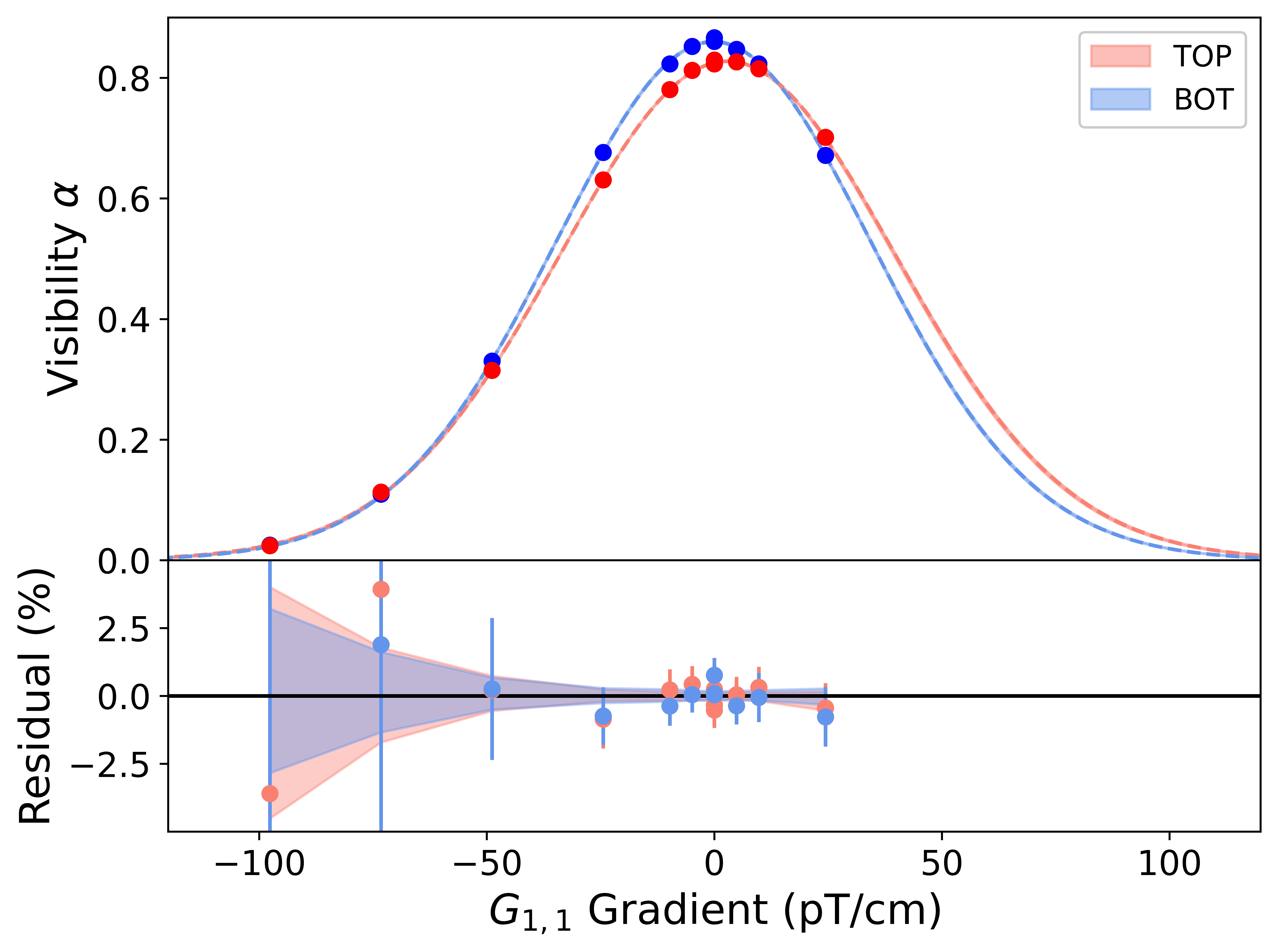}
\caption{Spin depolarization for different magnetic gradients in the top (red) and bottom (blue) storage chambers of n2EDM. The data points are the extracted visibility from measuring the asymmetry at each gradient (see Figure~\ref{fig:visibility}). The lower panel in each graph shows the percent residual between the data and theory. The error on the residual points is the normalized data statistical error, whereas the band shows the calculation $\pm1\sigma$ normalized spread from the bootstrapped minimization. (TOP): Gravitationally enhanced depolarization from the vertical gradient $G_{1,0}$. (BOTTOM): Intrinsic depolarization from linear gradient $G_{1,1}$. The fitted gradients $G_{1,-1},G_{2,0}$ are not shown here for brevity, but are in the supplemental and of similar quality.}
\label{fig:depolarization_data}
\end{figure}

\begin{figure}[t]
\includegraphics[width=0.90\columnwidth]{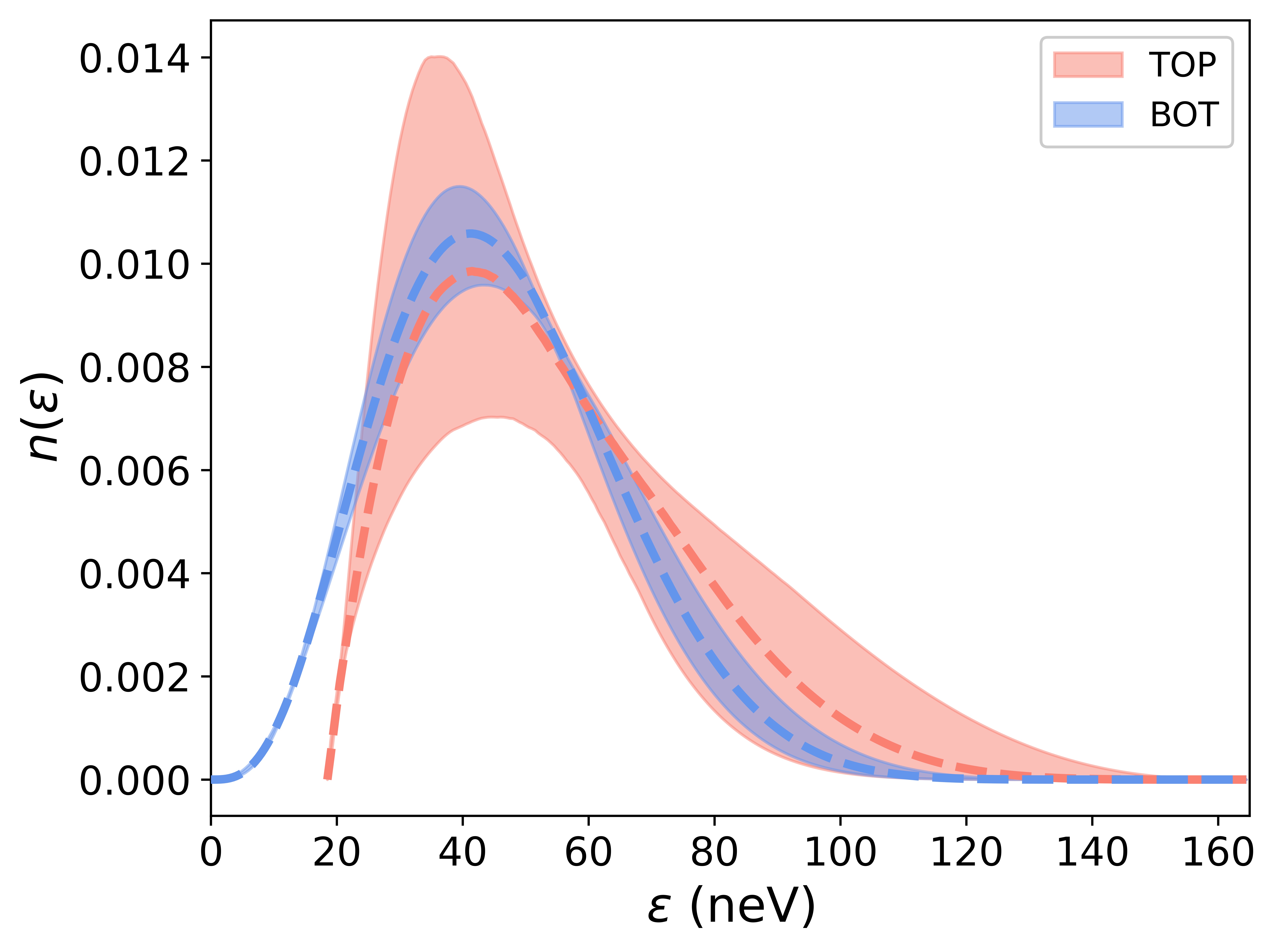}
\includegraphics[width=0.90\columnwidth]{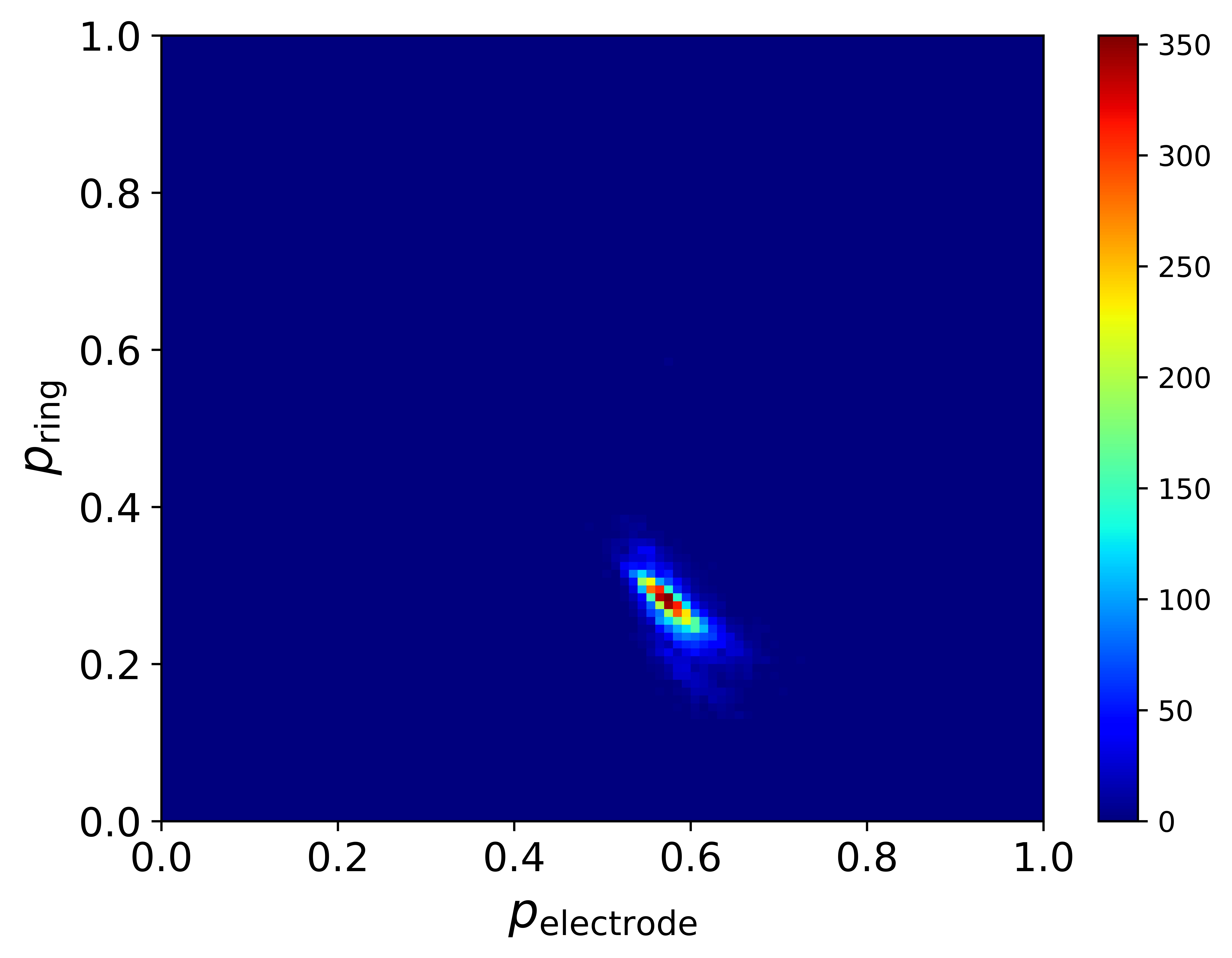}
\caption{(TOP): Extracted energy spectrum for the top (red) storage chamber and bottom (blue) storage chamber. The dashed line is the $50$th percentile and bands demonstrate $16-84$ percentile spread from the bootstrapped minimization. (BOTTOM): Correlation of the diffuse parameters for the electrode and ring surfaces of the bottom storage chamber. The color scale indicates the number of solutions from the bootstrapped minimization in a given bin. Here, $p=0$ would be perfectly specular reflections, while $p=1$ is perfectly diffuse. See text for discussion.}
\label{fig:results}
\end{figure}

We simultaneously fit four depolarization data sets (using Eqns.~\ref{eqn:grav_depolarization}-\ref{eqn:tau2mag}) for the two storage chambers separately: $\alpha(G_{1,0})$, $\alpha(G_{1,1})$,  $\alpha(G_{1,-1})$, $\alpha(G_{2,0})$. In each data set, we remove background contamination from the detected UCN counts (see online supplemental materials). We parameterize the energy spectrum as $n(\epsilon)=n_0(\epsilon-\epsilon_\textrm{min})^a(\epsilon_\textrm{max}-\epsilon)^b$, which ensures vanishing density below $\epsilon_\textrm{min}$ and above $\epsilon_\textrm{max}$ (such as the lowest Fermi potential of the storage chamber), and $n_0$ is a normalization such that $\int_{\epsilon_\textrm{min}}^{\epsilon_\textrm{max}}n(\epsilon)d\epsilon=1$. As mentioned, $\epsilon_\textrm{min}$ in the top chamber is expected to be shifted by roughly $18$~\si{\nano\electronvolt} due to the UCNs gaining energy as they fill into the chamber from the top. In the bottom, $\epsilon_\textrm{min}=0$. Due to the correlation between $\epsilon_\textrm{max}$ and $b$, we fix $\epsilon_\textrm{max}=165$~\si{\nano\electronvolt} to be sufficiently high (above the expected Fermi potential) and let $b$ control how quickly the energy spectrum decays. The extracted spectrum is independent of $\epsilon_\textrm{max}$ for sufficiently high values. 
We allow $\alpha_0$ to vary for each data set, and allow for an inherent gradient shift in each data set (i.e. an offset in $G$). 

 
Figure~\ref{fig:depolarization_data} shows the comparison between the data and the calculated theory from the minimization results. Remarkably, we can simultaneously fit all data to a few percent across extremely large gradients and across a large depolarization scale, from $\alpha\sim1$ down to $\alpha\sim0$. While $G_{1,0}$ exhibits a small systematic shifts at large gradients, the effect is only a few percent and is due to the coil inducing other gradients at such high currents. Figure~\ref{fig:results} shows the resulting energy spectrum for the individual storage chambers of n2EDM and the diffuse parameters for the electrodes and rings. Parameter correlations and central values for both chambers are given in the online supplemental material. 


Figure~\ref{fig:results} demonstrates the sensitivity of using spin depolarization to extract the energy spectrum and storage properties. We emphasize that UCNs with energies below $30$~\si{\nano\electronvolt} are most strongly affected by gravitational depolarization, providing the strictest constraints on $n(\epsilon)$. Since we require a continuous energy spectrum and normalization to $\alpha_0$, this constrains the high-energy tail. However, our results are consistent with different energy spectrum parameterizations (see online supplemental materials). The larger variance in the top chamber results from a lack of UCN statistics in the lowest energy bin, and therefore a larger space of $n(\epsilon)$ is possible. 

\begin{figure}[t!]
\includegraphics[width=0.90\columnwidth]{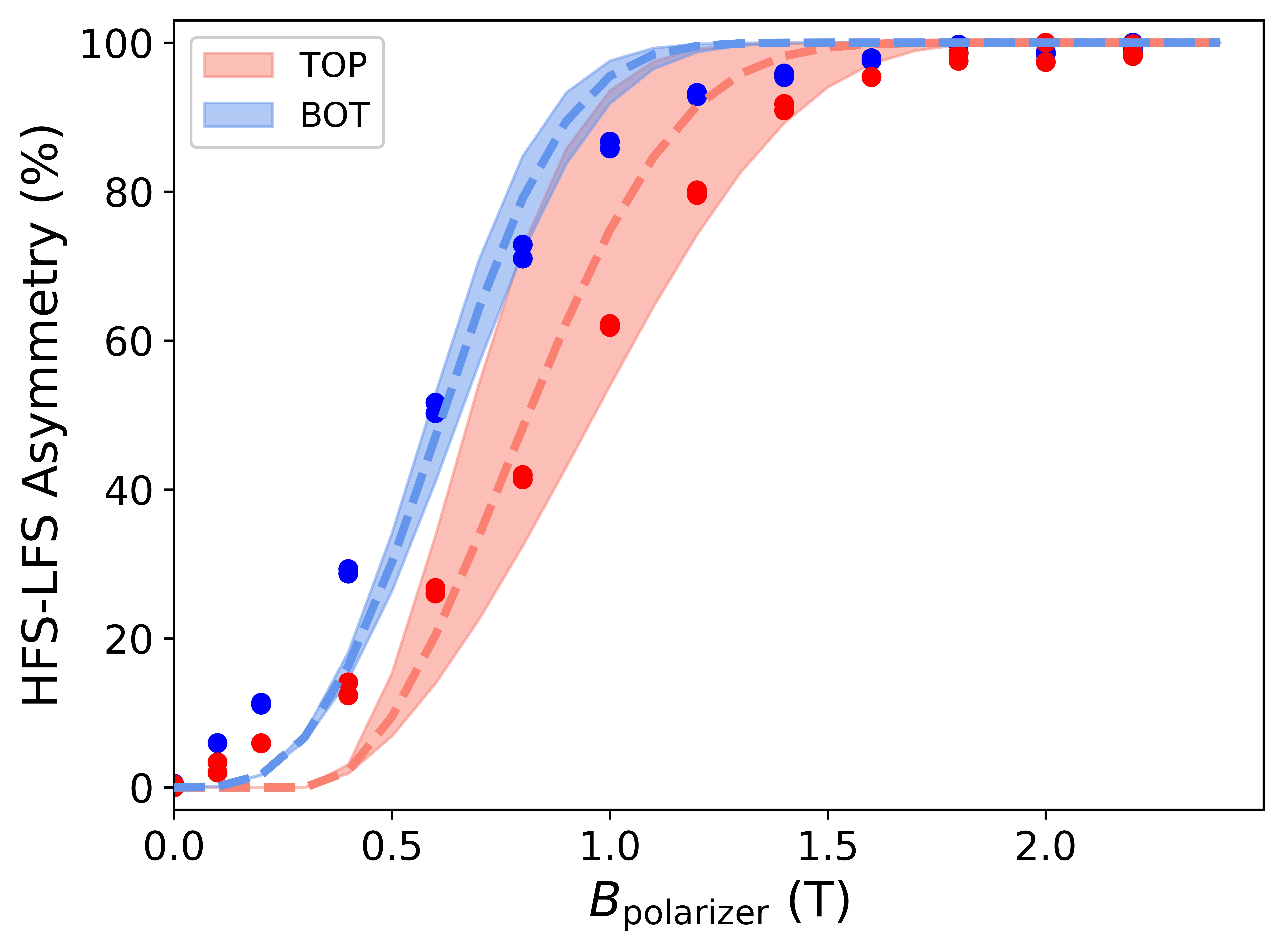}
\caption{\label{fig:scm_scan}LFS-HFS neutron asymmetry as a function of polarizing magnet strength $B_\mathrm{polarizer}$ for the top (red) chamber and bottom (blue) chamber. The points are the measured spin-asymmetry after 200~\si{\second} storage, rescaled to lie between 0-100\%. The curves are using the extracted energy spectra from Figure~\ref{fig:results}. The statistical uncertainty on the asymmetry is shown but not visible and on the order of $1\%$. See text for details.}
\end{figure}

Figure~\ref{fig:results} (bottom) also shows the sensitivity to the average diffuse parameters describing the surfaces of the storage chamber. We emphasize that the dependencies of $\alpha_\textrm{int}$ on $G_{1,1}$ and $G_{2,0}$ allow the disentanglement of $p_\mathrm{electrode}$ and $p_\mathrm{ring}$. This is due to the different types of trajectories that contribute to the correlation times $\tau_{1,1}$ and $\tau_{2,0}$ that sample the storage volume differently. As mentioned, the bottom chamber has the best sensitivity and tightly constrains $p_\mathrm{electrode}$ and $p_\mathrm{ring}$. The top chamber has a larger spread and is consistent (see the online supplemental materials). 

To validate our result on the energy spectrum, we compare to a measurement (not utilized in our fit) where neutrons up to different energies were polarized and then stored in our material trap.
Before the $5$~\si{\tesla} polarizing magnet of n2EDM, neutrons are in a superposition of spin states, and can be described as being in a ``low-field seeking'' (LFS) state or ``high-field seeking'' (HFS) state~\cite{PaulNStorage,HFSTrapPSI2011}. The LFS neutrons will be reflected depending on their energy and the strength of the polarizing field, whereas the HFS neutrons will always pass, increasing the spin-asymmetry with increasing polarizing field. 
This approach is also described in Ref.~\cite{losalamosucn}.

Using the energy spectrum of Figure~\ref{fig:results}, we calculate the LFS-HFS asymmetry as a function of polarizing field. Since the polarizing magnet will filter based on the longitudinal energy spectrum (not the total energy spectrum), we do a simple conversion that $E_\mathrm{long}=E_\mathrm{tot}\braket{\cos^2(\theta)}$, where we average over $\theta$, the azimuthal angle of neutrons along the beam-path guides using input from MCUCN~\cite{MCUCN,MCUCN_2}. Additionally, we must correct the total energy spectrum for the potential-energy difference at the polarizing magnet height compared to the storage chambers. Finally, we assume that the HFS spectrum at $5$~\si{\tesla} and LFS spectrum at $0$~\si{\tesla} are equivalent in the chamber. We compare our prediction with the measurements as described in Figure~\ref{fig:scm_scan}. The prediction agrees reasonably well, though not perfectly. This is largely due to the averaging to convert to longitudinal energy (see online supplemental materials). It could also be due to unmodeled energy-dependent or spin-dependent losses (such as the transmission asymmetry through the aluminum vacuum separator in the polarizing magnet~\cite{PSI_UCNStorage}), a variable efficiency of spin detection, or a breakdown of our assumption on the LFS spectrum.

\section{\label{sec:conclusion}Conclusions}
In this work, we described a new technique for leveraging spin depolarization mechanisms to extract UCN energy spectrum and, simultaneously, storage chamber properties. We emphasized the sensitivity of this result by comparing the two identical storage chambers at different heights and different orientations for filling of the n2EDM experiment~\cite{n2edmDesign2021}. We validated the robustness of this result by comparing to an independent measurement (using a magnetic filter) which is also sensitive to the energy spectrum. 

Lastly, we can use our work to predict an important systematic effect for n2EDM. n2EDM will use a $^{199}$Hg co-magnetometer to cancel temporal magnetic field drifts in the extracted UCN resonance frequency. However, there is a frequency shift of the neutron precession frequency relative to mercury in the presence of a vertical magnetic field gradient due to the gravitational offset of the slow neutrons. Using this work and the energy spectrum, we estimate the vertical offset of the UCN ensemble due to gravity:
\begin{eqnarray*}
    \braket{z}^\mathrm{TOP} &= -0.16(1)\mathrm{\ cm} \\
    \braket{z}^\mathrm{BOT} &= -0.22(1)\mathrm{\ cm}
\end{eqnarray*}
which can be used to quantify the size of the systematic effect. This prediction will be tested in an upcoming measurement using both UCNs and Hg-199, following Ref.~\cite{MagFieldUni2019}.
\section{Acknowledgments}
We appreciate the ongoing excellent technical support by Michael Meier.
Specific technical support by L. Noorda and N. Kohler is acknowledged. We acknowledge the excellent support by the
BSQ group running the UCN source, the 
HIPA accelerator operating crew, 
and various PSI support groups.

Support by the Swiss National Science Foundation Projects 188700 (PSI), 163413 (PSI), 178951 (PSI), 204118 (PSI), 172626 (PSI), 213222 (PSI), 212754 (PSI), 10001566 (PSI), 169596 (PSI), 181996 (Bern), 236419 (ETH), 200441 (ETH), 10003932 (ETH), and FLARE 186179, 201473, 216603, 232701 are gratefully acknowledged.
This project has received funding from the European Union’s Horizon 2020 research and innovation programme under the Marie Skłodowska-Curie grant agreement No 884104.
This work is support by the DFG (DE) by the funding of the PTB core facility center of ultra-low magnetic field KO 5321/3-1 and TR 408/11-1.
The LPC Caen and the LPSC Grenoble acknowledge the support of the French Agence Nationale de la Recherche (ANR) under reference ANR-14-CE33-0007 and the ERC project 716651-NEDM.
University of Bern acknowledges the support via the European Research Council under the ERC Grant Agreement No. 715031-Beam-EDM.
The Polish collaborators wish to acknowledge support from the National Science Center, Poland, under grant No. 2018/30/M/ST2/00319, and No. 2020/37/B/ST2/02349, and No. 2016/23/D/ST2/00715, as well as by the Minister of Education and Science under the agreement No. 2022/WK/07.
Support by the Cluster of Excellence ‘Precision Physics, Fundamental Interactions, and Structure of Matter’ (PRISMA+ EXC 2118/1) funded by the German Research Foundation (DFG) within the German Excellence Strategy (Project ID 39083149) is acknowledged.
Collaborators at the University of Sussex wish to acknowledge support from the School of Mathematical and Physical Sciences, as well as from the STFC under grant ST/S000798/1, ST/W000512/1, as well as from the STFC under grants ST/S000798/1 and ST/W000512/1.
This work was partly supported by the Fund for Scientific Research Flanders (FWO), and Project GOA/2010/10 of the KU Leuven.
Researchers from the University of Belgrade acknowledge institutional funding provided by the Institute of Physics Belgrade through a grant by the Ministry of Education, Science and Technological Development of the Republic of Serbia.
University of Kentucky also acknowledges support by US Departent of Energy under contract DE-SC0014622.



\bibliography{references}

\end{document}